%% file: GPD.tex
\documentclass{article}

\input{GPD-macros}

\begin{document}

\title{Testing Randomness by Matching Pennies}

\author{Dusko Pavlovic
	\hspace{3em} Peter-Michael Seidel \hspace{3em} Muzamil Yahia \\ University of Hawaii, Honolulu, USA\\
  \texttt{\{dusko,pseidel,muzamil\}@hawaii.edu}}
  
  \date{}

\maketitle

\begin{abstract}
In the game of Matching Pennies, Alice and Bob each hold a penny, and at every tick of the clock they  simultaneously display the head or the tail sides of their coins. If they both display the same side, then Alice wins Bob's penny; if they display different sides, then Bob wins Alice's penny. To avoid giving the opponent a chance to win, both players seem to have nothing else to do but to randomly play heads and tails with equal frequencies. However, while not losing in this game is easy, not missing an opportunity to win is not. Randomizing your own moves can be made easy. Recognizing when the opponent's moves are not random can be arbitrarily hard.

The notion of randomness is central in game theory, but it is usually taken for granted. The notion of outsmarting is not central in game theory, but it is central in the practice of gaming. We pursue the idea that these two notions can be usefully viewed as two sides of the same coin. The resulting analysis suggests that the methods for strategizing in gaming and security, and for randomizing in computation, can be leveraged against each other.

\end{abstract}

\section{Introduction}
\subsection{Game of Matching Pennies}
The payoff matrix for Matching Pennies is displayed on Table~\ref{fig-matching}. For the convenience of using the bitstring notations, we denote the heads move as 0 and the tails move as 1. 
The game is repeated, and we assume that it is played long enough that even the smallest strategic advantages are captured in the outcome. Both players can win or lose arbitrarily large amounts of pennies.
\vspace{-1\baselineskip}
\begin{table}[htbp]
\begin{center}
\begin{tabular}{r||lc|lc|}
\multicolumn{5}{r}  {0} \hspace{3.3em} {1}\ \ 
\\
\cline{2-5}
         &          & $-1$ &          & $1$ \\ 
         &&&& \\
{0} & $1$ &         & $-1$ &           \\
\cline{2-5}
         &          & $1$ &         & $-1$\\ 
         &&&& \\
{1}    & $-1$ &          & $1$       &   \\
\cline{2-5}
\end{tabular}

\caption{\small Payoffs for Matching Pennies}
\label{fig-matching}
\end{center}
\end{table}

\subsection{How not to lose Matching Pennies}\label{sec-NE}

To determine her strategy, Alice might reason something like this.
\begin{quote}
Suppose that I consistently play 1 with a frequency $p\in [0,1]$ and thus 0 with a frequency $1-p$. If I set $p\lt \frac 1 2$, then Bob can get the expected payoff $-p+(1-p) = 1-2p \gt 0$ by playing 1. If I set $p\gt \frac 1 2$, then Bob can get the expected payoff $p - (1-p) = 2p-1 \gt 0$ by playing 0. If I set $p = \frac 1 2$, then Bob's expected payoff is the same whether he plays 1 or 0: it is $1-2p = 2p -1 = 0$. Since Bob's winnings are my losses, the best strategy for me is to set $p=\frac 1 2$, and to play 0 and 1 with equal frequencies, since that minimizes my expected losses.
\end{quote}
By the same reasoning, Bob arrives at the same conclusion, that he should set the frequency of playing 1 at $q = \frac 1 2$. This is the well known Nash equilibrium of the game of Matching Pennies. Both players arrive to it by minimizing the expected losses.

\subsection{Playing Matching Pennies}
In general, a mixed strategy Nash equilibrium prescribes the frequencies for both players' moves in the long run. The essential assumption is that the moves will be \emph{randomized}. If Bob's move is predictable with some likelihood, then Alice can increase her chances to win. It seems natural to imagine that the players randomize by tossing their coins, and displaying the random outcomes. At the equilibrium, the players are just passive servants of chance, since they cannot gain anything by deviating from it. If they are rational, all they can do is toss their coins.

But suppose that Bob suddenly plays 
\beq\label{eq01} 0101010101010101010101010101010101010101\eeq 
Will Alice predict that Bob's next move is $0$ and play $0$ to win a penny? If she thinks probabilistically, she will probably notice that the probability of getting \eqref{eq01} by flipping a fair coin is $2^{-40}$, which is exactly the same as the chance of getting, e.g.
\beq\label{eqrand}
1101000100110101001011100100000100000010\eeq
or any other sequence that she would accept as random.  
If Alice's rationality is based on probabilities, then she will not be able to distinguish any two strings of Bob's moves, since they are all equally probable if he tosses fair coins.

But if Bob knows, or even just believes, that Alice's rationality is based on probabilities, and that Alice will thus continue to randomize her moves in any case, then Bob has no reason to randomize, since playing  \eqref{eq01}, or \eqref{eqrand}, or a string of 0s, or any other string, yield the same expected payoff against Alice's random plays. On the other hand, if Alice believes that Bob's rationality is based on probabilities, then she will have no reason to randomize either, for the same reason as Bob. 

So by combining their beliefs about their probabilistic reasoning, both players will become indifferent towards mixing and randomizing their moves. Their common knowledge that they may both stop randomizing, because they both know that the opponent will be unable to tell, will not change their expected payoffs. Indeed, if they both play non-randomly, one of them will almost surely win and the other will lose, but their chances to be the winner are the same, and they average out. However, while the expected payoffs remain unchanged, the higher moments will, of course, cha\-nge significantly.

\subsection{How to win Matching Pennies if you can}
In order to exploit Bob's deviation from the equilibrium, or to give him an incentive to genuinely randomize his mixed equilibrium strategy, Alice must go beyond probabilities, i.e. beyond just calculating the frequency of his moves. If she just checks whether the frequencies of 0 and 1 are $\frac 1 2$, she will detect that the string consisting of 0s alone is not random, but not that the string \eqref{eq01} is not random; if she checks whether the frequencies of 00, 01, 10 and 11 are $\frac 1 4$, she will detect that \eqref{eq01} is not random, but not that the string where these four digraphs of bits alternate is not random; etc. By checking that each bitstring of length $n$ has \emph{in the long run}\/ the frequency $\frac 1 {2^n}$, she will detect many non-random plays, but still miss most of them. E.g., the string 
\beq\label{eq123} 
011011100101110111100010011010101111001101\ldots
\eeq 
obtained by concatenating the binary notations for the sequence of natural numbers 0,1,2,3\ldots will pass the bias tests for all $n$-grams, if taken long enough, yet it is, of course, easily predictable, and obviously not random. Moreover, Bob might, e.g., randomize all even bits, and just alternate 0s and 1s at the odd positions. To recognize such opportunities, Alice will have to check that \emph{every substring}\/ of the string of Bob's past moves has unbiased frequencies of all $n$-grams. As the game goes on, Alice will thus have to keep proving that Bob's play, i.e. the ever growing string of his past moves, is what von Mises called \emph{Kollektiv}\/ in his theory of probability \cite{vonMises}. Proving that something is a \emph{Kollektiv}\/ is known to be a problematic task, as specifying the substrings to be tested has led to problems that remained open for many years \cite{Ville:collectif,Bienvenu-Shafer-Shen:martingales}.

\subsection{Randomness from equilibrium}

Scratching the surface of the basic assumption about the players' incentive to implement a mixed strategy equilibrium led us straight into the foundations of probability. There is, of course, nothing surprising about the fact that the concept of a mixed strategy, expressed in terms of probability, depends on the foundations of probability. The point is not so much that there are deep foundational problems lurking behind simple games. It seems much more useful, and more interesting, that, the other way around, there seem to be instructive ways to state the solutions of the foundational problems of probability in terms of games.

In particular, we show that the usual definition of mixed strategy equilibria based on the notion of randomness as given can be reversed, and that the notion of randomness can be defined using mixed strategy equilibria. The upshot is not just that a complicated concept of randomness is replaced by an intuitive game theoretic concept of not losing Matching Pennies at the equilibrium; the upshot is also that the effective content of both concepts, of randomness and of equilibrium, can be analyzed in terms of \emph{computational power of testing}. This formalization brings both the basic probabilistic concepts and the basic game theoretic concepts into the logical realm of computable inductive inference \cite{FisherR:1959,SolomonoffR:64,Blum-Blum:inductive,RissanenJ:book,wallace2005statistical}. 

\subsection{Background and related work}\label{sec-related}
We propose a simple and narrow bridge between games and probabilities. An extensive effort towards reconstructing the foundations of probability theory from a particular game has been ongoing for many years, as reported by Shafer and Vovk \cite{Shafer-Vovk:probability}. The work presented in this paper is not only at the opposite end of the scale in terms of its scope and technical sophistication, but it also goes in a different direction, and therefore uses an essentially different model. While the authors of \cite{Shafer-Vovk:probability} aim to reconstitute the full power of the diverse probabilistic tools in their rich gaming model, the point here is to illustrate how the most basic games capture the most basic probability concepts in a natural fashion. A similar analysis geared in the opposite direction of \emph{eliminating}\/ probabilities is provided in \cite{
PavlovicD:AMAI17}.

The bridge between games and probabilities is built using significance testing and computation. Significance testing goes back to Fisher \cite{FisherR:1959,FisherR:1959} and lies, of course, at the core of the method of statistical induction. The constructions sketched here are related to the computational versions of testing, developed on one hand in Martin-L\"of's work \cite{Martin-Loef:randomness,NiesA:book}, and on the other hand in the techniques of inductive learning \cite{Gold67,Blum-Blum:inductive,Vovk:Learning}. 

We analyze the computational content of testing. The analyses of the computational content of strategic reasoning go back to the earliest days of game theory \cite{RabinMO:effective-strat}, and continue through theory of bounded rationality \cite{RubinsteinA:bounded}, and on a broad front of algorithmic game theory \cite{NisanN:AGT}. The finite state machine model seems preferred for specifying strategies \cite{RubinsteinA:bounded,halpern-pass-automata}, since computable strategies lead to problems with the equilibrium constructions \cite{Knoblauch,Nachbar-Zame}. In recent work, a different family of problems, arising from the cost of strategic computations has been analyzed, including the cost of randomization \cite{halpern-pass-cost,halpern-pass-sequential}. This led the players to not just lose the incentive to randomize, as in the little story above, but to prefer determinism. Although we are here also looking at the problem of  deviating from the equilibrium into non-ran\-dom\-ness, we are concerned with a completely different question: \emph{How should the opponent recognize and exploit this deviation?} The present work seems to deviate from previous computational approaches to gaming in one essential aspect: we are not analyzing the computations that the players perform to construct or implement their own strategies, or the equilibrium, but the computations that they perform to test the opponents' strategies. This leads into a completely different realm of computability, that emerges from a different aspect of gaming. While the analysis goes through for most models of computation, represented by an abstract family of pro\-gram\-ma\-ble functions, as explained in Sec~\ref{sec-programmable}, it is perhaps worth stating the obvious: that stronger notions of computation lead to stronger notions of randomness. 

Although the recently introduced high level models of gaming \cite{Abramsky-Winschel:MSCS,PavlovicD:CALCO09,Hedges-Oliva-Winschel} are not explicitly introduced in the paper, as they are not necessary for the presented results, they were used in the original versions of the presented results, and can perhaps be recognized in the background.

\subsection{Outline of the paper}
In Sec.~\ref{sec-prelim} we spell out the preliminaries and some notations used in the paper. In Sec.~\ref{sec-testing} we motivate and explain the simplest case of randomness testing, with respect to the uniform distributions, and describe its application in gaming. 
Sec.~\ref{sec-randomness} derives as a corrolary the characterization of random strings as the equilibrium plays. In Sec.~\ref{sec-gen-testing} we describe how to construct randomness tests for arbitrary pro\-gram\-ma\-ble distributions. Sec.~\ref{sec-conclusions} closes the paper with some final comments.

\section{Notations and preliminaries} \label{sec-prelim}

\subsection{Monoid of plays} 

In the games considered in this paper, the set of \emph{moves}\/ is always $\TTwo = \{0,1\}$. We sometimes call 0 heads and 1 tails. A \emph{play}\/ is a finite string (or list, or vector) of moves $\vec x = x_1 x_2 x_3\cdots x_m$, or $\vec y = y_1 y_2 y_3\cdots y_n$ played in a match of a game.  The set of all bitstrings, used to represent plays, is denoted by $\Plays$. The empty bitstring is $()$, and the concatenation of bitstrings is $\cons{\vec x} {\vec y} = x_1\cdots x_my_1\cdots y_n$. They constitute the monoid $\big(\Plays, ::, ()\big)$, freely generated by $\TTwo$. The monoid structure induces the prefix ordering
\bea\label{eq-prefix}
\vec x \sqsubseteq \vec y & \iff & \exists \vec z.\ \cons {\vec x} {\vec z} = \vec y
\eea
and the \emph{length} measure $\ell :\Plays \to \NNn$, which is the unique homomorphism from the free monoid over two generators to the free monoid over one generator. The fact that the length measure is a homomorphism means that
\[ \ell() = 0 \qquad \mbox{and} \qquad \ell\left(\cons {\vec x}{\vec y}\right) = \ell\left(\vec x\right) + \ell\left(\vec y\right) \]
We shall also need a bijective pairing $<-,->: \Plays \times \Plays \to \Plays$ with the projections $-_{(0)}, -_{(1)}: \Plays\to \Plays$, which means that together they satisfy
\bear
\left<\vec x_{(0)}, \vec x_{(1)}\right> = \vec x & \quad & \left<\vec x_{0}, \vec x_{1}\right>_{(i)} = \vec x_i
\eear
Using the fact that a free monoid is also cofree, a bijective pairing can be derived from any two disjoint injections $\Plays\inclusion \Plays$.  For simplicity, we use
\bea\label{eq-def-pairing}
<\vec x,\vec y> & = & x_1 x_1 x_2 x_2\cdots x_m x_m 01 y_1 y_2\cdots y_n 
\eea
where $\vec x = x_1x_2\cdots x_m$ and $\vec y = y_1 y_2\cdots y_n$. The length induces the shift homomorphism
\bea\label{eq-pairing}
\ell\left(<\vec x, \vec y>\right) & = & 2\ell(\vec x) + \ell(\vec y) + 2
\eea

\subsection{Sets and functions}

 $|X|$ denotes the number of elements of the set $X$. A function written $f:X\rightarrow Y$ is always total, whereas a partial function is written $h: X\pfn Y$. We write $h(x)\halts$ when the partial function $h$ is defined on the input $x$, and $h(x)\nohalts$ or $h(x) = \uparrow$ when $h$ is undefined on $x$. 

\subsection{Programmable functions}\label{sec-programmable}
We say that $f:\Plays \pfn \Plays$ is \emph{$\LLL$-pro\-gram\-ma\-ble}, or that it is an \emph{$\LLL$-function}\/ when it is specified using a programming language $\LLL$. The intuitions from the reader's favorite programming language, practical or theoretical, should do. For a theoretical example, one could take $\LLL$ to be the language of finite state machines. A program could then be either a list of transitions of a Moore or Mealy machine, or a corresponding regular expression   \cite{Hopcroft-Ullman,ConwayJH:regular}. The graphs of pro\-gram\-ma\-ble functions would be regular as languages.  A larger family or pro\-gram\-ma\-ble functions would be obtained from a Turing complete programming language, like Python or Java, or from the language of Turing machines themselves. In the latter case, a program could again be a list of the transitions of the machine. A high-level formalism is based on the structure of \emph{monoidal computer}, spelled out in \cite{PavlovicD:IC12,PavlovicD:MonCom3
}. 

Formally, the programming language is given by a \emph{universal evaluator}\/ (or \emph{interpreter}), a partial function $\interpreter : \Plays \times \Plays \pfn \Plays$. This function may or may not be in $\LLL$. E.g., when $\LLL$ is a Turing complete language, then then $\interpreter$ is an $\LLL$-function. If $\LLL$ is the language of regular expressions, then their universal evaluator is not $\LLL$-pro\-gram\-ma\-ble. 

We usually write $\LLL(x,y)$ in the form $\uev x y$. A universal evaluator is characterized by the requirement that for every $\LLL$-function $f:\Plays  \to \Plays$ there is a bitstring  $p_f: \in \Plays$ such that 
\bear
f(\vec x)\  & = &\ \ \  \uev{p_f} \vec x 
\eear

\section{Randomness for uniform distributions}\label{sec-testing}
We focus on Alice's task to detect patterns of non-ran\-dom\-ness in Bob's play, which she could exploit to predict his moves. Bob is assumed to be doing the same, observing Alice's play and trying to detect some patterns. But what is a pattern? And what does it mean to detect it?

Intuitively, an object has a pattern if it can be described succinctly, i.e. compressed. E.g. the string in \eqref{eq01} can be compressed to $(01)^{20}$ in mathematical notation, or to  
{\footnotesize 
\begin{lstlisting}
for (i=0; i<20; i++) { print 01 }
\end{lstlisting}}
\noindent in a Java-like programming language. The program to extend \eqref{eq01} infinitely would be $(01)^\ast$ or 
{\footnotesize 
\begin{lstlisting}
for (;;) { print 01 }
\end{lstlisting}}
\noindent and the program to output \eqref{eq123} would be
{\footnotesize
\begin{lstlisting}
for (i=0;;i++) { print i }
\end{lstlisting}}
\noindent On the other hand, a program to output the string \eqref{eqrand}, without a detectable pattern, would have to spell it out in full length:
{\footnotesize
\begin{lstlisting}
print 110100010011010100101110010000010000001001
\end{lstlisting}}
The idea that randomness can be defined as incompressibility goes back to Kolmogorov \cite{Kolmogorov:Sankhya}, and further back to the scholastic logical principle known as \emph{Occam's Razor}, which established the priority of succinct descriptions as inductive hypotheses, as explained by Solomonoff \cite{SolomonoffR:64}.

\subsection{Testing \detectors}
\medskip
\be{defn}\label{def-detector}
Let $\LLL$ be a family of pro\-gram\-ma\-ble (partial) functions. A \emph{\detector}\/ is an $\LLL$-function $h: \Plays \pfn \Plays$ such that 
\bea\label{eq-decompression}
h(\vec x) = \vec y  & \Longrightarrow & \ell(\vec x)\lt \ell(\vec y)
\eea
A string $\vec y$ that lies in the image of $h$ is said to be \emph{$h$-regular}. A string $\vec x$ on which $h$ is defined and maps it to $\vec y$ is a \emph{short description}\/ of $\vec y$.
%
A \detector\  $h$ is \emph{predictive}\/ if 
\bea\label{eq-predictive}
\forall \vec x.\ h(\vec x)\halts &\Longrightarrow & \exists \vec z. \ \vec x \sqsubset \vec z \wedge h(\vec x) \sqsubset h(\vec z)
\eea
where $\sqsubset$ is the prefix ordering \eqref{eq-prefix}.
\ee{defn}

 The tacit idea behind predictive hypotheses is that the input data are given with some end markings, which tell the computer where the input string ends. This is the case with the data input on most real computers, but not on \emph{"plain"}\/ Turing machines, which leads to the restriction to \emph{prefix-free}\/ or \emph{self-delimiting}\/ machines \cite{Levin-Zvonkin,Vitanyi:book,Downey:book}. For the Turing machine model, the reader should assume that there is a special symbol $\square$ denoting the end of each string, and that the string inclusion ignores that symbol.  The computation $h\left(\vec x\right)$ thus halts when it encounters $\square$ after $\vec x$, whereas the computation $h\left(\vec z\right)$ proceeds longer and provides a longer output when $h$ is predictive. 

\be{defn} 
A bitstring $\vec y$ is said to be \emph{$h$-regular at the level $m\in \NNn$}\/ if
\bea 
\exists \vec x.\ \ \ h(\vec x) = \vec y  & \wedge & \ell(\vec x) +m \leq \ell(\vec y)
\eea
The $h$-regular bitstrings at each level form the \emph{$h$-regularity sets} 
\bea
H_m^n & = & \left\{ \vec y \in \TTwo^n | \exists \vec x.\ h(\vec x) = \vec y \wedge \ell(\vec x) +m \leq \ell(\vec y)\right\} \label{eq-Hmn}\\
H_m & = &  \bigcup_{n=1}^\infty H_m^n \label{eq-Hm}
\eea
Setting for convenience $H_0 = \Plays$ yields a decreasing sequence of sets:
\beq \label{eq-tower}
H_0 \supseteq H_1\supseteq H_2 \supseteq H_3 \supseteq \cdots \supseteq H_m\supseteq \cdots \eeq
This tower of sets is the \emph{$h$-test}.
\ee{defn}

Note that a bitstring of length $n$ can only be regular at the level $m$ if $m\leq n$. The $h$-regularity sets $H^n_m$ for $m\gt n$ are empty.

\be{prop}\label{prop-shrink} The size of $h$-regularity sets decreases exponentially with $m$, in the sense 
\bea\label{eq-shrink}
\lvert H^n_m\rvert & \lt & 2^{1+n-m}
\eea
\ee{prop}

\begin{proof}
By \eqref{eq-Hmn}, for every $\vec y \in H^n_m$ there is $\vec x$ such that $h(\vec x) = \vec y$ and $\ell(\vec x) + m \leq \ell(\vec y)$, and thus $\ell(\vec x) \leq n-m$, because $\vec y \in \TTwo^n$. The function $h:\Plays \to \Plays$ is thus restricted to a surjection  onto $H^n_m$ from the set of strings $\vec x$ of lengths at most $n-m$. Hence \eqref{eq-shrink}.
\end{proof}

Proposition~\ref{prop-shrink} says that the chance that an observation $\vec y$ is $h$-regular at the level $m$ decreases exponentially in $m$. Since this is true for all \detectors, the implication is that most bitstrings are irregular: most \detectors\  are eventually rejected, and most bitstrings are accepted as random.  This is a formal expression of Laplace's observation that regular objects constitute a null set \cite{Laplace}.  

\be{defn}\label{def-significance} The \emph{$h$-regularity degree}\/ $\Sig_h(\vec y)$ is the highest $h$-re\-gu\-la\-ri\-ty level that the bitstring $\vec y$ achieves, i.e.
\bear
\Sig_h(\vec y) & = & \max \{m\leq \ell(\vec y)\ |\ \vec y\in H_m\} 
\eear 
The $h$-regularity degree is thus a function $\Sig_h:\Plays \to \NNn$.
\ee{defn}

\subsection{Alice's testing strategy}\label{sec-hyp-testing}
Alice's computations of $h$-regularity degree follow the basic method of significance testing \cite{FisherR:1925,FisherR:1959}. She tests whether Bob's play $\vec y$ satisfies the \detector\  $h$. The hypothesis is rejected if $\vec y$ is not $h$-regular at a sufficiently high level. So Alice goes down the test tower $H_0\supseteq H_1\supseteq H_2 \supseteq \cdots$, and checks how far is it true that $\vec y \in H_m$. This ceases to be true when  $m = \Sig_h(\vec y)$. The \detector\ $h$ is thus rejected if the regularity degree $\Sig_h(\vec y)$ is below some \emph{significance threshold}\/ $M$, chosen in advance.  If she wants to echo statisticians' habit to set the significance level at $1\%$ or $5\%$, Alice should probably choose $M$ to be between 4 and 7, since the indices of the test towers correspond to the negative logarithms of statistical significance levels.  

But what is Alice trying to achieve by testing Bob? What will she do if she detects a significant $h$-regularity in his play $\vec y$? She wants to predict his moves, and use the prediction to take his pennies. In particular, if she finds a significantly shorter description $\vec x$ of $\vec y$ realized by $h$, she will try to guess a bitstring $\vec s$ such that $h\left(\cons {\vec x} {\vec s}\right)$ is defined, and extends $\vec y$, i.e. such that \bear 
\vec y & \sqsubset &  h\left(\cons {\vec x} {\vec s}\right)
\eear
The definition of \emph{predictive}\/ \detectors\ requires that they always allow such extensions. So if she formulates a predictive \detector\ $h$, finds a short description $\vec x$ of Bob's play $\vec y$, and guesses an extension $\vec s$ allowing her to predict Bob's moves, Alice will match and take Bob's pennies.

\subsection{Separating regularity and randomness}\label{sec-separating}
The essence of Alice's testing strategy is to separate a regular component of Bob's strategy from the random component. If Bob plays completely randomly, his play $\vec y$ will not have a short description, and Alice will not find a \detector\ $h$ that $\vec y$ will satisfy. The regular component is then empty. If Alice finds a \detector\ $h$ and a short description $\vec x$ of $\vec y$ proving that $\vec y$ satisfies $h$, then $h$ captures some of the regularity of $\vec y$. If there is still some regularity in $\vec x$, then it has a still shorter description $\vec x'$, realized using a \detector\ $h'$. In other words, $\vec x = h'\left(\vec x'\right)$ and there is $m'$ such that $\ell\left(\vec x'\right) + m' \lt \ell\left(\vec x\right)$. But this means that $\vec x'$ is a still shorter description of Bob's play $\vec y$, showing that it satisfies the \detector\ $h\circ h'$ at the regularity level $m+m'$, since $h\circ h'\left(\vec x'\right) = \vec y$ and $\ell\left(\vec x'\right) + m+m'\lt \ell\left(\vec y\right)$. 

On the other hand, if $\vec x$ is incompressible, then it is random. In that case, the short description $\vec x$ is the random component of Bob's play $\vec y$, whereas all of its regularity is captured by $h$. Alice can thus extrapoloate Bob's future moves by running $h$. She also has to expand the random part $\vec x$ by some additional randomness $\vec s$, as presumably Bob will continue doing. In that sense, Alice still has to gamble. But just like $\vec x$ is much shorter than $h\left(\vec x\right) = \vec y$, the chance of guessing $\cons{\vec x}{\vec s}$ is greater than the chance of guessing $h\left(\cons{\vec x}{\vec s}\right)$. So separating out the regular component $h$ of Bob's play and reducing the randomness of Bob's play to a significantly shorter description presents a significant advantage for Alice.

Since $\Sig_h\left(h\left(\vec x\right)\right)\geq \ell\left(\vec x\right)$, regularity increases with length, and the testing outcomes become more significant, and provide better fitting predictions. On the other hand, longer strings also fit more \detectors, and the usual problems of \emph{overfitting}\/ in statistical inference enter scene. But testing \detectors\ as $\LLL$-pro\-gram\-ma\-ble functions turns out to have a special feature, which we consider next. 

\subsection{Universal \detector}\label{sec-universal}
The main remaining question is: \emph{How should Alice choose her \detectors?} 
She can, of course, stares at $\vec y$ and search for a pattern. She can try a \detector\  $h^{(1)}$, and if it gets rejected, she can try $h^{(2)}$, and $h^{(3)}$, and so on. But which one should she try first?  Occam suggests: \emph{The simplest \detectors\ should be tried first}. But which ones are the simplest? Solomonoff and Kolmogorov suggest: \emph{The simplest functions are those that have the shortest programs}\/  \cite{Vitanyi:book,RissanenJ:book,wallace2005statistical}.

This is where Alice comes to use the fact that her \detectors\ are pro\-gram\-ma\-ble. By enumerating all programs, she can in principle test all \detectors. If the universal evaluator $\LLL$ is  $\LLL$-pro\-gram\-ma\-ble itself, she can in fact test a \emph{universal \detector}. The idea of a universal randomness test goes back to Per Martin-L\"of \cite{Martin-Loef:randomness}. Since it will eventually detect any regularity, any universal \detector\ test is in fact also a universal randomness test, as random strings can be characterized as just those that pass all tests \cite{vonMises}. 

\be{defn}\label{def-universal}
A \detector\  $\upsilon: \Plays \pfn \Plays$ is \emph{universal}\/ if any string that is regular with respect to any \detector\  $h$ is also regular with respect to $\upsilon$. More precisely, for every \detector\  $h: \Plays \pfn \Plays$ there is a constant $c_h$ such that for every bitstring $\vec x$ holds
\bea\label{eq-universal}
\Sig_h(\vec y) &\leq & c_h + \Sig_\upsilon(\vec y)
\eea
\ee{defn}

\be{prop}\label{prop-universal}
If the universal evaluator of a family of $\LLL$-pro\-gram\-ma\-ble functions is $\LLL$-pro\-gram\-ma\-ble itself, then the family also contains a universal \detector. \ee{prop}

The assumption that the universal evaluator $\LLL$ is $\LLL$-pro\-gram\-ma\-ble is satisfied not just when $\LLL$ interprets a Turing complete language, but also when it is resticted to a complexity class with complete instances.

\be{proof}
Let $\uev{}: \Plays \times \Plays \pfn \Plays$ be a universal evaluator. Recall that this means that for every computable function $f: \Plays \pfn \Plays$ there is a program $\vec p_f$ such that $f(\vec x)  =  \uev{\vec p_f} \vec x$. 
Define
\beq\label{eq:uni-detect}
\begin{split}
\upsilon(\vec x) \  = \  \begin{cases}
\uev{\vec x_{(0)}}\vec x_{(1)} & \mbox{ if }\ell(\vec x) \lt 
\ell\left( \uev{\vec x_{(0)}}\vec x_{(1)}\right)
\\
\uparrow & \mbox{ otherwise}
\end{cases}
\end{split}
\eeq
Then $\upsilon: \Plays \pfn \Plays$ is a \detector\  by Def.~\ref{def-detector}.  For any \detector\  $h: \Plays \pfn \Plays$ and any bitstring $\vec x$ then holds
\bear
 h(\vec x) & = & \upsilon\left(<\vec p_h, \vec x>\right)
\eear
where $\vec p_h$ is a program encoding $h$.  Then for $c_h = 2\ell\left(\vec p_h\right)+2$ holds
\bea\label{eq-down}
c_h + \ell(\vec x) + m \leq \ell\left(\upsilon\left(<\vec p_h, \vec x>\right)\right) = \ell\left(h(\vec x)\right)
\eea
were the constants 2 come from the particular definition of pairing and length that we have chosen in \eqref{eq-def-pairing} and \eqref{eq-pairing}. But \eqref{eq-down} means that  $\upsilon$-regularity at the level $m$ implies $h$-regularity at the level $m+c_h$, i.e. $U_m^n \subseteq H_{m+c_h}^n$. Hence \eqref{eq-universal}.
\ee{proof}

\subsection{Alice's universal strategy}\label{sec-uni-testing}
If Bob's play $\vec y$ is not random, then finding an $\vec x$ such that $\upsilon\left(\vec x\right) = \vec y$ will separate the regular component $\vec x_{(0)}$ and the random component $\vec x_{(1)}$ from his play $\vec y$, as explained in Sec.~\ref{sec-separating}. Guessing Bob's moves $\vec b$ can then be reduced to the task of guessing a shorter random string $\vec s$ such that $\cons{\vec y}{\vec b} = \upsilon\left(\cons{\vec x}{\vec s}\right)$.

In summary, Alice's tasks are similar to her testing strategy: the \textbf{first task}\/ is to search for inverse images along a programmable function, this time $\upsilon : \Plays \to \Plays$; and her \textbf{second task} is again to use the detected regularity of $\vec y$ to predict an extension.

%
%
%
%
%
%

Concerning the first task, note that Alice will stall if she simply lists a sequence of candidates $\vec x^{(1)}, \vec x^{(2)}, \vec x^{(3)},\ldots$ and tries to compute $\upsilon\left(\vec x^{(i)}\right) = \uev{\vec x^{(i)}_{(0)}}\vec x^{(i)}_{(1)}$ for $i = 1,2,3,\ldots$, one after another seeking to find an inverse image of $\vec y$ along $\upsilon$. That strategy will only go as far as the first $\vec x^{(i)}$ for which the program $\vec x^{(i)}_0$ diverges on the input $\vec x^{(i)}_1$; the next candidate will never be tested. To avoid that, the search for short descriptions $\vec x$ must  proceed by \emph{dovetailing}, as described e.g. in \cite{Levin-Zvonkin}. This means that the search through the sequence $\vec x^{(1)}, \vec x^{(2)}, \vec x^{(3)},\ldots$ should run a finite number of steps of each computation in a finite initial segment of the sequence, and keep extending that initial segment. In that way, each member of the sequence will eventually be reached and run. E.g., a single step of each of the computations $\upsilon\left(\vec x^{(i)}\right)$ can be run in following order:
$$ \vec x^{(1)}, \vec x^{(2)}, \vec x^{(1)}, \vec x^{(2)}, \vec x^{(3)},  \vec x^{(1)}, \vec x^{(2)}, \vec x^{(3)}, \vec x^{(4)}, \vec x^{(1)}, \ldots$$

Once Bob's play $\vec y$ has been captured by a short description $\vec x$, i.e. decomposed in the form $\vec y = \uev{\vec x_{(0)}} \vec x_{(1)}$, where $\vec x_{(0)}$ is the regular component of $\vec y$, and $\vec x_{(1)}$ is its random component, then Alice can proceed with the second task. 

In summary, Alice's universal strategy can be described as the search for the earliest bitstring $\vec b$ which results from a shorest extension $\vec s$ of a shortest inverse image $\vec x$ of $\vec y$ along $\upsilon$. This can be summarized as the function $\alice: \Plays \pfn \Plays$ where
\bear
\alice (\vec y) & = & \mu \vec b.\ \Alice\left(\vec y, \vec b\right)\\
&& \mbox{where}\\
\Alice\left(\vec y, \vec b\right) & \iff & \exists \vec x \vec s.\  \widetilde{\Alice}\left(\vec x, \vec y, \vec s, \vec b\right)  \ \wedge \\
&& \forall \vec x' \vec s'.\  \Big(\widetilde{\Alice}\left(\vec x', \vec y, \vec s', \vec b\right)\ \Rightarrow \\
&& \ell\left(\vec x\right)\leq \ell\left(\vec x'\right)  \wedge \ell\left(\vec s\right)\leq \ell\left(\vec s'\right)\Big) 
\\ && \mbox{where}\\
\widetilde{\Alice}\left(\vec x, \vec y, \vec s, \vec b\right) & \iff & \upsilon\left(\vec x\right) = \vec y \ \wedge\ \uev{\vec x_{(0)}}\left(\cons{\vec x_{(1)}}{\vec s}\right) = \cons{\vec y}{\vec b}
\eear
While Alice thus seeks to predict Bob's moves $\vec b$ in order to play the same moves, Bob's universal strategy would be dual, in the sense that he would seek to predict Alice's moves $\vec a$ in order to play the opposite moves. We discuss below what happens if two universal strategies are played against each other.

\section{Matching Pennies randomness}\label{sec-randomness}

The notion of randomness as incompressibility, as formalized by Kolmogorov \cite{Kolmogorov:Sankhya} and developed in algorithmic information theory \cite{Vitanyi:book}, has been justified by Martin-L\"of's proof that incompressible strings are just those that pass all randomness tests \cite{Martin-Loef:randomness,NiesA:book,Vitanyi:book}. But we have seen that randomness tests are also a part of playing Matching Pennies. The players stay at the equilibrium only as long as their plays pass each other's tests. Whenever a test produces a significant outcome, the randomness hypothesis is rejected, and the players depart from the equilibrium, whether the detected pattern was a real consequence of someone's earlier deviation from the equilibrium, or whether the test overfitted a pattern onto an actually random string. The equilibrium persists only if both players' plays pass both players' tests. 

\be{corollary}\label{prop-randomness}
A bitstring is uniformly random (in the sense of Kolmogorov \cite{Kolmogorov:Sankhya,Martin-Loef:randomness}) if and only if it can occur as a play of the equilibrium strategy in the game of Matching Pennies.
\ee{corollary}

\bprf{ idea}
If a bitstring is uniformly random, then it will pass every randomness test, and can occur as an equilibrium strategy. If a bitstring can occur in an equilibrium strategy, and thus passes every randomness test, then it is uniformly random.
\epr

The upshot of this corollary is that randomness tests are an important aspect of the actual process of gaming, yet they are generally abstracted away from game theory. When randomness is taken for granted, the computational content of equilibrium constructions are abstracted away from game theoretic analyses, while the competitive aspects of gaming, of course, essentially depend on using randomness, and recognizing non-randomness. Taking the randomness \emph{testing}\/ for granted hides from sight the whole wide area of players' strategic analyses of each other's plays, which is where the essence of real gaming is played out. If Alice's play passes Bob's tests, but Alice's tests detect the regularity behind Bob's play, then Alice will win by outsmarting Bob. Randomness and outsmarting are two sides of the same coin. Taking one for granted hides the other one from sight, and separates game theory from practice.

While the concept of randomness in the above statements largely follows the approach and the ideas of Martin-L\"of \cite{Martin-Loef:randomness,Downey:book,NiesA:book,Vitanyi:book}, the abstract view of computation \cite{PavlovicD:IC12,PavlovicD:MonCom3
}, although lurking in the background in this extended abstract, allows a broader approach. When $\LLL$ is a Turing complete language, and testing is computable, then Prop.~\ref{prop-universal} implies that there is a universal strategy, and Prop.~\ref{prop-randomness} thus says that  a bitstring is uniformly random if it does not lose the game of Matching Pennies against the universal strategy. Using weaker programming languages $\LLL$, and thus specifying weaker randomness tests, yields weaker notions of randomness. A path towards a taxonomy of different notions of randomness obtained in this way is discussed  in \cite{NiesA:book}. The point here is that all such notions can be cast in terms of games. A different approach to a similar idea has been pursued to a much greater depth in \cite{Shafer-Vovk:probability}.

\section{Randomness for general distributions}\label{sec-gen-testing}
Equilibrium strategies for the game of Matching Pennies are mixed uniformly, in the sense that each move is assigned equal probability. Other games require other mixtures, with different probabilities assigned to different moves. It is easy to show that, any probability distribution over a finite set can arise as a mixed strategy equilibrium for a game with suitable payoffs. Moreover, iterated games, and games with dynamically changing payoffs, induce in the same way the various forms of probability distributions over sets of strings of  fixed lengths. To study dynamics, and some particular stochastic processes that arise in computation and cryptography, such probability distributions are often bundled together into a particular kind of measures, which we call string distributions. They also arise in a general form of mixed strategies for iterated games, and for games with dynamically changing payoffs. And while such mixed strategies require randomizing according to string distributions, their randomness can in turn be defined in terms of mixed strategy equilibria again.

\subsection{Testing $P$-\detectors\  and $P$-randomness}

\begin{defn}
A \emph{string distribution} is an $\LLL$-pro\-gram\-ma\-ble\footnote{The programmability of a real function $P$ can be defined in different ways. The idea going back to Turing \cite{TuringA:Entscheidung} is to present $P$ as a program $\vec p_P : \Plays \to \Plays$ which for each $\vec x$ outputs a program $\vec p_P(\vec x)$ which streams the digits of the real number $P(\vec x) \in [0,1]$.} function $P:\Plays \to [0,1]$ such that
\[P() = 1 \qquad \qquad P(\vec x) = P(\cons{\vec x}0) + P(\cons{\vec x} 1)
\] 
\ee{defn}

\begin{defn}
Given a string distribution $P:\Plays \to [0,1]$, a $P$-\detector\ with respect to $\LLL$ is an $\LLL$-function $h_P:\Plays \pfn \Plays$ such that
\bea\label{eq-p-detector}
h_P(\vec x) = \vec y &\Longrightarrow & \ell(\vec x) \lt \ell(\vec y)\  \wedge \ P(\vec x) \gt P(\vec y)
\eea
%
A bitstring $\vec y$ is said to be \emph{$h_p$-regular}\/ at the level $m\in \NNn$ whenever
\begin{multline} 
\exists \vec x.\ \ \ h_P(\vec x) = \vec y\ \wedge 
\ \ell(\vec x)+m\leq \ell(\vec y) \ \wedge\  P(\vec x)\geq 2^m \cdot P(\vec y)
\end{multline}
The $h_P$-regular bitstrings at each level form the \emph{$h_P$-regularity sets} 
\begin{multline}
H_m^n  \  = \ \big\{ \vec y \in \TTwo^n\ |\ \exists \vec x.\ h_P(\vec x) = \vec y\  \wedge \\ \wedge\ \ell(\vec x)+m\leq \ell(\vec y) \ \wedge\   P(\vec x)\geq 2^m\cdot P(\vec y)\big\} \label{eq-pHmn}
\end{multline}
The sets $H_m = \bigcup_{n=1}^\infty H_m^n$ form the $h_p$-test
$$H_1\supseteq H_2 \supseteq H_3 \supseteq \cdots \supseteq H_m\supseteq \cdots$$
\ee{defn}

\be{prop}\label{prop-p-shrink} The $P$-size of $h_P$-regularity sets decreases exponentially with $m$
\bea\label{eq-pshrink}
\sum_{\vec y\in H^n_m} P(\vec y) & \lt & 2^{1-m}
\eea
\ee{prop}

\begin{proof}
By definition of $H^n_m$, for every $\vec y\in H^n_m$ there is $\vec x$ of length at most $n-m$ such that $P(\vec y) \leq 2^{-m} P(\vec x)$. It follows that
\[\sum_{\vec y\in H^n_m} P(\vec y) \leq \sum_{\vec x\in \TTwo^{n-m}} 2^{-m}\cdot P(\vec x)\ \lt\ 2^{1-m} \hspace{2em}
\]
\end{proof}

The search for non-random patterns, deviating from a given string distribution $P$, proceeds just like the search for patterns deviating from the uniform distribution in Sec.~\ref{sec-hyp-testing}. When $\LLL$ is the family of all computable functions, there is a universal $P$-\detector, defined just like in Def.~\ref{def-universal}.


\be{prop}\label{prop-universal-P}
If the universal evaluator $\LLL$ and the string distribution $P$ are $\LLL$-pro\-gram\-ma\-ble, then there is an $\LLL$-pro\-gram\-ma\-ble universal $P$-\detector\  as well. \ee{prop}

\bpr
The universal  $P$-\detector\  is this time
\beq\label{eq:uni-detect-P}
\begin{split}
\upsilon_P(\vec x) \  = \  \begin{cases}
\uev{\vec x_{(0)}}\vec x_{(1)} & \mbox{ if }\ \ \ \ \ \ell(\vec x) \lt 
\ell\left( \uev{\vec x_{(0)}}\vec x_{(1)}\right) 
\\
& \mbox{ and }\ P(\vec x) \gt 
2^{2\ell\left(\vec x_{(0)}\right)+2}\cdot P\left( \uev{\vec x_{(0)}}\vec x_{(1)}\right) 
\\
\uparrow & \mbox{ otherwise}
\end{cases}
\end{split}
\eeq
were the 2s come again from \eqref{eq-def-pairing} and \eqref{eq-pairing}. Setting $c_{h} = 2\ell\left(\vec p_{h}\right)+2$ again where $\vec p_{h}$ is a program for $h_P$ we have
\begin{gather*} 
c_{h}  + m +  \ell(\vec x)\ \ \leq\ \ \ell\left(\upsilon_P\left(<\vec p_{h}, \vec x>\right)\right)\ \ =\ \ \ell\left(h_P(\vec x)\right)\\
P(\vec x)\  \  \gt\ \  2^{m}\cdot P\left(\upsilon\left(<\vec p_{h}, \vec x>\right)\right)\ \ \geq\   \ 2^{c_h+m}  \cdot P\left(h_P(\vec x)\right)
\end{gather*}
which gives $U_m^n \subseteq H_{c_h + m}^n$. By Def.~\ref{def-universal}, this means that $\upsilon_P$ is universal for all $P$-\detectors\  $h_P$.
\epr

\be{corollary}\label{prop-randomness-P}
A bitstring is $P$-random if and only if it can occur as a play in a game where the string distribution $P$ is a component of a mixed strategy equilibrium. 
\ee{corollary}

\subsection{Universal strategies beyond Matching Pennies}\label{sec-uni-strategy}
Just like a universal \detector\  can be used to build a universal strategy for winning, if possible, in the game of Matching Pennies, universal $P$-\detectors\ can be used to build universal strategies for a large family of games, where mixed strategies are expressed in terms of string distributions. A familiar example of such a game is the iterated version of Prisoners' Dilemma, where dynamically changing mixed strategies, inducing string distributions, have been played in tournaments against each other since the early days. Strategies played in such tournaments are usually described by finite state machines, and thus induce $\LLL$-programmable distributions where $\LLL$ is a language generating regular expressions. More powerful languages allow specifying not only more powerful strategies, but also games where the payoff matrices are not necessarily fixed through the iterations of the game, but may also change, in an $\LLL$-programmable way. The crucial feature that allows analyzing such games are the fixed point constructions, enabled by the universal evaluators.

We assume that the payoffs are public information, and that both Alice and Bob have both computed the equilibrium strategies, and know the distributions $P_A$ and $P_B$ according to which Alice and Bob, respectively, must randomly mix their moves. Alice's first task is thus to program a function $\eta_A :\Plays \to \Plays$ to search for short descriptions $\vec x = \eta_A\left(\vec y\right)$ of Bob's plays $\vec y$, whereas Bob's first task is to program a function $\eta_B :\Plays \to \Plays$ to search for short descriptions $\vec u = \eta_B\left(\vec w\right)$ of Alice's plays $\vec w$. So they are both looking for a right inverse of the universal $P$-detector of the opponent's string distribution $P$, i.e. 
\[
\upsilon_B \circ \eta_A\left(\vec y\right) = \vec y \qquad \qquad \upsilon_A \circ \eta_B \left(\vec w\right)\  = \  \vec w 
\]
where we write $\upsilon_B$ to simplify $\upsilon_{P_B}$ and $\upsilon_A$ for $\upsilon_{P_A}$. Their second tasks will be to program functions $\vartheta_A, \vartheta_B:\Plays \to \Plays$ to guess the likely extensions of opponents' plays, i.e.
\[
\upsilon_B \circ \vartheta_A \circ \eta_A\left(\vec y\right) \sqsupset \vec y \qquad \qquad \upsilon_A \circ \vartheta_B \circ \eta_B \left(\vec w\right)\  \sqsupset \  \vec w 
\]
In summary, Alice's and Bob's tasks are thus to program strategies $\alice = \vartheta_A \circ \eta_A$ and $\bobb = \vartheta_B \circ \eta_B$ with
\[
\upsilon_B \circ \alice \left(\vec y\right) \sqsupset \vec y \qquad \qquad \upsilon_A \circ \bobb \left(\vec w\right)\  \sqsupset \  \vec w 
\]
Alice's universal strategy described in Sec.~\ref{sec-uni-testing} is an instance of this $\alice$.

\section{Concluding remarks}\label{sec-conclusions}

The starting point of this paper was the observation that finding and playing one's own strategy is often much easier than recognizing and understanding other players' strategies. In particular, randomizing is much easier than testing randomness. On the other hand, knowing that the opponent keeps an eye on how you play is necessary for the implementations of many equilibrium concepts, usually assumed implicitly. In order to stay at an equilibrium, the players must test each other. But capturing their tests opens an alley towards modeling competition, outsmarting, and deceit, which are prominent in the practice of gaming, but often ignored in game theory. We believe that the tools are readily available to tackle this interesting and important aspect of gaming. 

Players' randomness testing of each other's plays turned out to be an intuitive characterization of the notion of randomness. It is perhaps worth emphasizing  here that the players with different computational powers recognize different notions of randomness. More precisely, different families of pro\-gram\-ma\-ble functions $\LLL$ induce different \detectors, different tests, different notions of randomness, and different implementations for the mixed strategy equilibria. Restricting the family $\LLL$ to the language of regular expressions or finite state machines would give a weak but interesting notion. The \detectors\  could be implemented along the lines of the familiar compression algorithms, such as those due to Lempel, Ziv and Welch \cite{Ziv-Lempel78,Welch84}. However, since there is no such thing as a universal finite state machine, capable of evaluating all finite state machines, such tests based on regular languages would have to be specified one at a time, and sought ad hoc. In contrast, taking $\LLL$ to be a Turing complete language, such as the language of Turing machines themselves, allows constructing a universal randomness test, which Alice could implement as a universal \detector\ from Sec.~\ref{sec-universal}. This leads to the canonical notion of randomness spelled out by Kolmogorov, Martin-L\"of and Solovay, and characterized in Corollary~\ref{prop-randomness}. Although the simple dovetailing technique used to construct the universal \detector\ quickly leads beyond the realm of what is considered feasible computation,  the methods of algorithmic learning and statistical induction  are built upon them  nevertheless \cite{Hutter:UAI,RissanenJ:book,wallace2005statistical}. 
%
%

\bibliographystyle{plain}
\bibliography{GPD-refs,PavlovicD,games,logic}  
\end{document}

%% file: GPD-macros.tex
\usepackage{etex}
\usepackage{graphicx}
\usepackage{pgf}
\usepackage{color}
\usepackage{bbm}
\usepackage{amsmath,amsthm,amsfonts,amssymb}

  \usepackage{hyphenat} 
\usepackage{pstricks}
\usepackage{rotating}
\usepackage{enumerate}

\usepackage{listings}

{\end{itemize}}

{\end{itemize}\vspace{-.6\baselineskip}}

{\end{enumerate}}

\renewcommand{\paragraph}[1]{\ \ {\sc #1}}

\renewcommand{\to}{\longrightarrow}

\newcommand{\inclusion}{\hookrightarrow}

\newcommand{\pfn}{\rightharpoondown}

\newcommand{\halts}{\!\downarrow}
\newcommand{\nohalts}{\!\!\uparrow}

\newcommand{\cons}[2]{{#1}::{#2}}
\newcommand{\detector}{hypothesis}
\newcommand{\detectors}{hypotheses}
\newcommand{\Plays}{\TTwo^\ast}
\newcommand{\interpreter}{\LLL}

\newcommand{\uev}[1]{\mbox{\large$\left\{\mbox{\normalsize $#1$}\right\}$}}

\newcommand{\LLL}{{\cal L}}

\renewcommand{\Bbb}{\mathbb}

\newcommand{\NNn}{{\Bbb N}}

\newcommand{\TTwo}{\mathbbm{2}}

\mathcode`\<="4268 
\mathcode`\>="5269 
\mathchardef\gt="313E 
\mathchardef\lt="313C 

\newcommand{\be}[1]{\begin{#1}}

\newcommand{\ee}[1]{\end{#1}}
\newcommand{\beq}{\begin{equation}}
\newcommand{\eeq}{\end{equation}}
\newcommand{\ba}[1]{\begin{array}{#1}}
\newcommand{\ea}{\end{array}}
\newcommand{\bea}{\begin{eqnarray}}
\newcommand{\eea}{\end{eqnarray}}
\newcommand{\bear}{\begin{eqnarray*}}
\newcommand{\eear}{\end{eqnarray*}}
\newcommand{\bpr}{\begin{prf}{}}
\newcommand{\epr}{\end{prf}}
\newcommand{\bprf}[1]{\begin{prf}{#1}}
\newcommand{\eprf}{\end{prf}}

 \def\pushright#1{{
    \parfillskip=0pt            
    \widowpenalty=10000         
    \displaywidowpenalty=10000  
    \finalhyphendemerits=0      
    \leavevmode                 
    \unskip                     
    \nobreak                    
    \hfil                       
    \penalty50                  
    \hskip.2em                  
    \null                       
    \hfill                      
    {#1}                        
    \par}}                      

 \def\qed{\pushright{$\square$}\penalty-700 \smallskip}

\newenvironment{prf}[1]{\begin{trivlist} \item[{\sc ~Proof #1}.]}%
{\qed\end{trivlist}}

\newcommand{\Sig}{\sigma}

\newtheorem{thm}{Theorem}[section]
\newtheorem{defn}[thm]{Definition}

\newcommand{\alice}{\alpha}
\newcommand{\bobb}{\beta}
\newcommand{\Alice}{A}